\documentclass[conference]{IEEEtran}
\IEEEoverridecommandlockouts

\usepackage{cite}

\usepackage{color,soul}
\usepackage{subfig}
\usepackage{amsmath}
\usepackage{gensymb}
\usepackage{dsfont}
\usepackage{array}
\usepackage{booktabs}
\usepackage{filecontents}

\ifCLASSINFOpdf
  \usepackage[pdftex]{graphicx}
  \graphicspath{{../pdf/}{../jpeg/}}
  \DeclareGraphicsExtensions{.pdf,.jpeg,.png}
\else
  \usepackage[dvips]{graphicx}
  \graphicspath{{../eps/}}
  \DeclareGraphicsExtensions{.eps}
\fi

\usepackage{amsmath,amssymb,amsfonts}
\usepackage{algorithmic}
\usepackage{graphicx}
\usepackage{textcomp}
\usepackage{xcolor}
\usepackage{bm}
\usepackage[hyphens]{url}
\usepackage{lipsum}
\usepackage{float}
\usepackage[flushleft]{threeparttable}

\def\BibTeX{{\rm B\kern-.05em{\sc i\kern-.025em b}\kern-.08em
    T\kern-.1667em\lower.7ex\hbox{E}\kern-.125emX}}
\begin{document}
\bibliographystyle{ieeetr}

\title{A Time-Series Distribution Test System Based on Real Utility Data}
\author{\IEEEauthorblockN{Fankun Bu, Yuxuan Yuan, Zhaoyu Wang, Kaveh Dehghanpour, and Anne Kimber
\IEEEauthorblockA{Department of Electrical and Computer Engineering \\
Iowa State University\\
Ames, Iowa 50011 \\
fbu@iastate.edu, yuanyx@iastate.edu, wzy@iastate.edu, kavehd@iastate.edu, and akimber@iastate.edu\\}
}
\thanks{This work is supported by the National Science Foundation under CMMI 1745451 and the Department of Energy Office of Electricity under DE-OE0000875.

F. Bu, Y. Yuan, Z. Wang, K. Dehghanpour, and A. Kimber are with the Department of
Electrical and Computer Engineering, Iowa State University, Ames,
IA 50011 USA (e-mail: fbu@iastate.edu; wzy@iastate.edu)}
}

\makeatletter
\def\ps@IEEEtitlepagestyle{
  \def\@oddfoot{\mycopyrightnotice}
  \def\@evenfoot{}
}
\def\mycopyrightnotice{
  {\footnotesize
  \begin{minipage}{\textwidth}
  978-1-7281-0407-2/19/{\$31.00 \textcircled{c}2019 IEEE}
  \end{minipage}
  }
}

\maketitle
\begin{abstract}
In this paper, we provide a time-series distribution test system. This test system is a fully observable distribution grid in Midwest U.S. with smart meters (SM) installed at all end users. Our goal is to share a real U.S. distribution grid model without modification. This grid model is comprehensive and representative since it consists of both overhead lines and underground cables, and it has standard distribution grid components such as capacitor banks, line switches, substation transformers with load tap changer and secondary distribution transformers. An important uniqueness of this grid model is it has one-year smart meter measurements at all nodes, thus bridging the gap between existing test feeders and quasi-static time-series based distribution system analysis.
\end{abstract}


\section{Introduction}

Distribution test feeders have been designed to address various analytic challenges. The previous test feeders can be roughly classified into two categories based on their authenticity (Table \ref{tbl:feeder_feature}): \textit{Class I - Realistic Test Systems}: Making modifications on real distribution system, a number of standard test feeders have been developed to represent different types of networks \cite{WHK1991,WHK2010,RFA2010,KPS2014}. In \cite{WHK1991}, the Distribution System Analysis (DSA) subcommittee published the first five IEEE distribution test feeders, which are widely used among researchers. In \cite{WHK2010}, a comprehensive test feeder was proposed. Here ``comprehensiveness" means it contains all standard distribution grid components such as voltage regulator, distribution transformer and line switch. In \cite{RFA2010}, the IEEE 8500-Node Test Feeder was published to represent a large-sized unbalanced radial distribution system at both medium voltage (MV) and low voltage (LV) levels. In \cite{KPS2014}, the IEEE 342-Node Low Voltage Networked Test System with heavily meshed topology was developed to provide a benchmark for non-radial distribution network analysis. In \cite{NEV_test_case}, the IEEE Neutral-Earth-Voltage (NEV) Test Feeder was developed to study neutral conductor's voltage rise considering the resistance between the neutral conductor and earth ground. In \cite{ELVS}, the IEEE European Low Voltage Test Feeder was published for researchers to study typical low voltage networks in Europe. In \cite{EPRI_J1, EPRI_K1, EPRI_M1, EPRI_CK5ttoCKt24}, six real large-scale test feeders were published by Electric Power Research Institute (EPRI) to provide models for researchers interested in solar integration studies. \textit{Class II - Synthetic Test Systems}: Due to the limited number of standard test feeders, synthetic test systems have been developed as alternatives to flexibly represent various real networks. In \cite{PNNL_test_feeder}, 24 synthetic test feeders were presented, which characterize distribution systems in different regions of U.S. A clustering algorithm was used for developing these synthetic test feeders based on the data from 575 real distribution networks.

\begin{table*}[hbtp]
\begin{center}
	\renewcommand{\arraystretch}{1.35}
	\caption{Summary of Test Systems Features}\label{tbl:feeder_feature}
	\label{table1}
	\begin{tabular}{cccccc} \hline
		\textbf{Test System}  & \begin{tabular}[c]{@{}c@{}}  \textbf{Nominal}  \textbf{Voltage (kV)} \end{tabular} & \begin{tabular}[c]{@{}c@{}}  \textbf{Radial} \textbf{or Meshed} \end{tabular} & \begin{tabular}[c]{@{}c@{}}  \textbf{Feeder}   \textbf{Type} \end{tabular} & \begin{tabular}[c]{@{}c@{}}  \textbf{Distribution} \\ \textbf{Transformer} \end{tabular} & \begin{tabular}[c]{@{}c@{}}   \textbf{Real Time-series}\\ \textbf{Load Data}  \end{tabular}  \\
	    \hline
        IEEE 4 Node \cite{WHK1991} & 12.47, 4.16 or 24.9 & Radial & Four-wire wye & N  & N\\
        IEEE 13 Node \cite{WHK1991} & 115, 4.16, 0.48 & Radial & Four-wire wye & N  & N \\ IEEE 34 Node \cite{WHK1991} & 69, 24.9, 4.16 & Radial & Four-wire wye & N & N \\
	    IEEE 37 Node \cite{WHK1991} & 230, 4.8, 0.48 & Radial & Three-wire delta & N & N \\
    	IEEE 123 Node \cite{WHK1991} & 115, 34.5, 4.16, 0.48 & Radial & Four-wire wye & N & N \\
    	IEEE Comprehensive \cite{WHK2010} & 115, 24.9, 0.48, 0.24 & Radial & Four-wire wye & Y & N \\
    	IEEE 8500 Node \cite{RFA2010} & 115, 12.47, 0.24 & Radial & Four-wire wye & Y & N\\
    	IEEE NEV \cite{NEV_test_case} & 12.47, 0.24 & Radial & Four-wire wye & Y & N\\
    	IEEE 342 Node \cite{KPS2014} & 230, 13.2, 0.48, 0.24 & Meshed & Three-wire delta & Y & N \\
    	IEEE European LV \cite{ELVS} & 11, 0.416 & Radial & Four-wire wye & Y & Y, one day\\
    	EPRI J1 \cite{EPRI_J1} & 69, 12.47, 0.48, 0.24  & Radial & Four-wire wye & Y & N\\
    	EPRI K1 \cite{EPRI_K1} & 69, 13.2, 0.48, 0.24 & Radial & Four-wire wye & Y & N\\
    	EPRI M1 \cite{EPRI_M1} & 69, 12.47, 0.48, 0.24 & Radial & Four-wire wye & Y & N\\ EPRI Ckt5 \cite{EPRI_CK5ttoCKt24} & 115, 12.47, 0.48, 0.24 & Radial & Four-wire wye & Y & N\\
    	EPRI Ckt7 \cite{EPRI_CK5ttoCKt24} & 115, 12.47, 0.48, 0.24  & Radial & Four-wire wye & Y & N\\
    	EPRI Ckt24 \cite{EPRI_CK5ttoCKt24} & 230, 34.5, 13.2 0.48, 0.24  & Radial & Four-wire wye & Y & N\\
    	\hline
	\end{tabular}
\end{center}
\end{table*}

These existing works provide researchers useful benchmarks to utilize realistic distribution system models for power flow analysis, optimal equipment placement, islanded operation, renewable integration studies, state estimation and optimal power flow \cite{KPS2018}. However, Class I test feeders are stationary single-snapshot models, which generally lack real time-series measurements. This hinders their application for quasi-static time-series distribution system analysis, which requires capturing time-varying load behaviors \cite{KPS2018}. Furthermore, the outcomes of different systemic studies obtained from artificially-developed Class II test systems might not be easily generalized to practical networks. To the best of our knowledge, the IEEE European Low Voltage Test Feeder \cite{ELVS} is the only existing test feeder with time-series load data, as shown in Table \ref{tbl:feeder_feature}. Nonetheless, utilizing the European test feeder is still challenging for quasi-static time-series analysis due to three limitations: the first problem is the limited data length, which is confined to one day. This short period might not be sufficient for accurately tracking the time-varying behavior of the system in the long run. The second limitation is the absence of common network components in this test system, such as shunt capacitor banks and voltage regulators. The last limitation is that there exist differences in the distribution system structure and operation between the U.S. and Europe \cite{KPS2018}. Hence, it is necessary to publish test systems with long-term time-series load data, common network components, and real circuit models \cite{KPS2018}.

This paper presents a test system with one-year time-series power consumption data from more than 1120 customers.  Specifically, the test system is a real fully observable distribution network located in the Midwest U.S. The electric components include overhead lines and underground cables with a variety of phase configurations, load tap changing transformers, various secondary distribution transformers, shunt capacitor banks, and circuit breakers. The data includes one-year smart meter measurements of all customers, system component parameters, and detailed network topology. Our goal is to release a real distribution grid model and field measurements with minimum modification. However, when sharing the real data for research purpose and data analysis, it is important to consider data privacy issues \cite{RM2013}. To protect sensitive information of data owners, one minor modification is that the individual customers' power consumption (measured by smart meters) has been aggregated at the secondary distribution transformer level (120/240V). In this way, we avoid disclosing individual customers' load behaviors. The detailed information of the presented test system can be downloaded from our website \cite{zhaoyu_wang}.

\begin{figure*}[tbhp]
      \centering
      \includegraphics[width=2\columnwidth]{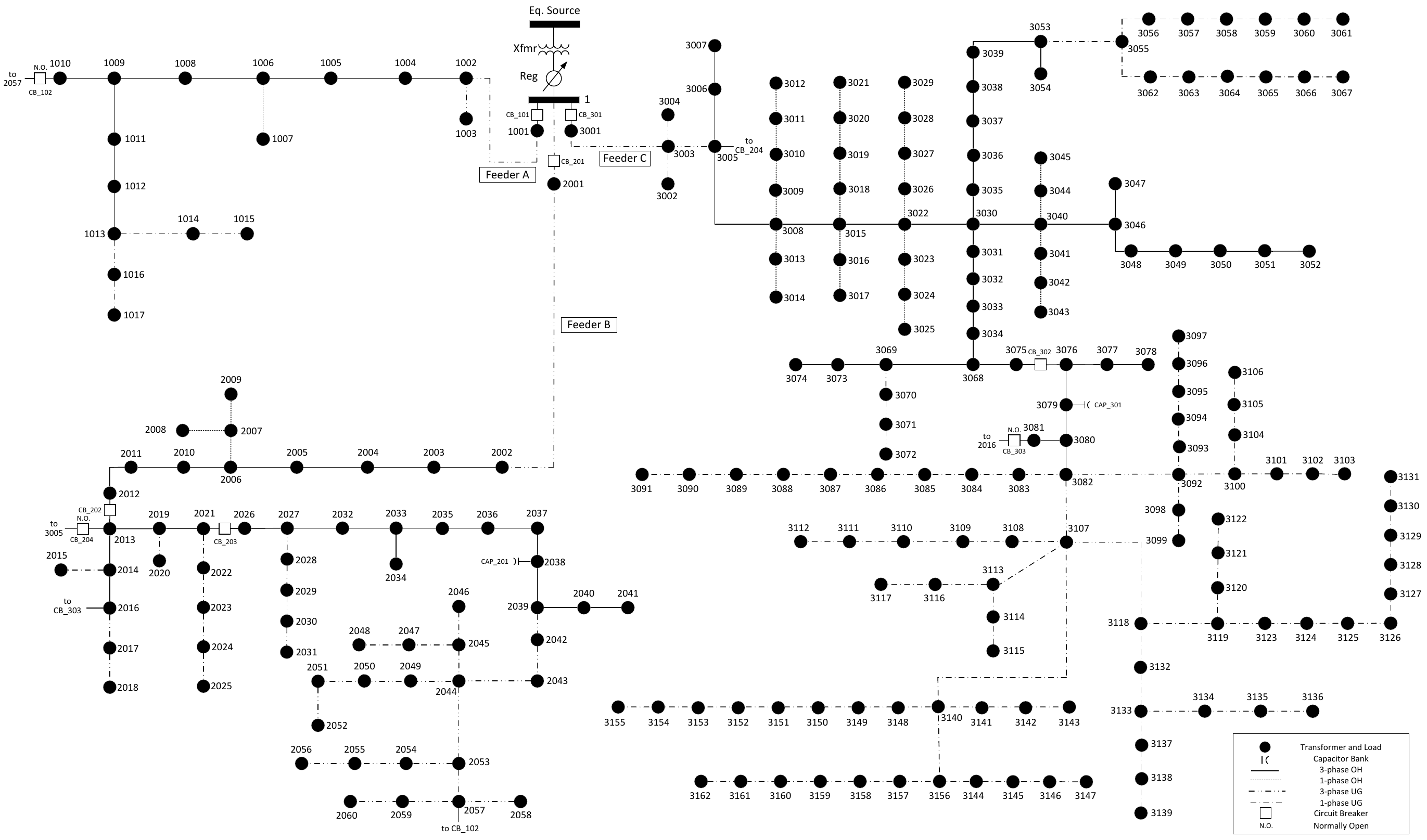}
\caption{One-line diagram of the test system.}
\label{fig:overall}
\end{figure*}
\begin{figure}
	\centering
	\includegraphics[width=3.2in]{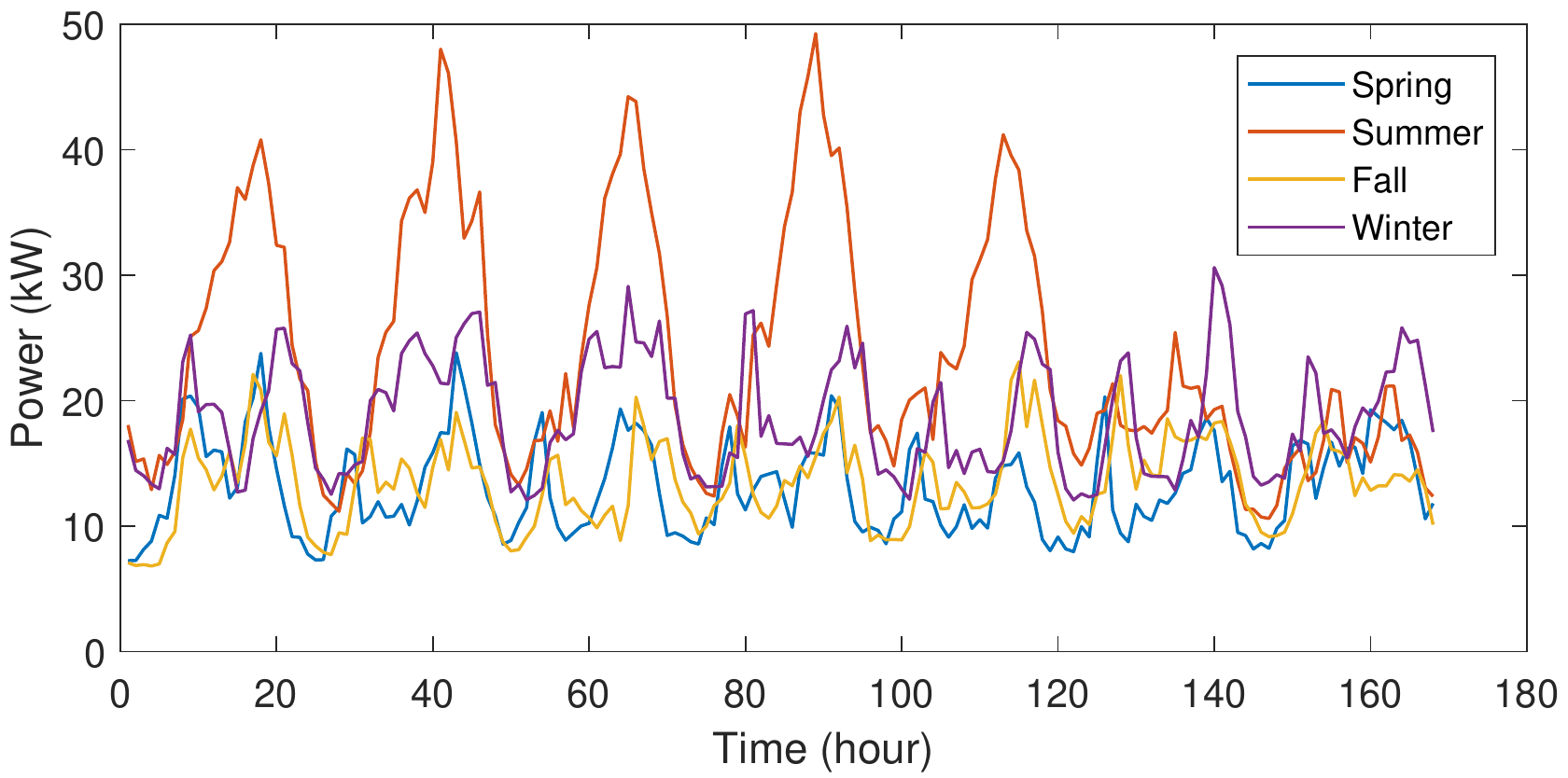}
	\caption{One week load profiles of a selected primary node in different seasons.}
	\label{fig:season}
\end{figure}
\begin{figure}
	\centering
	\includegraphics[width=2.5in]{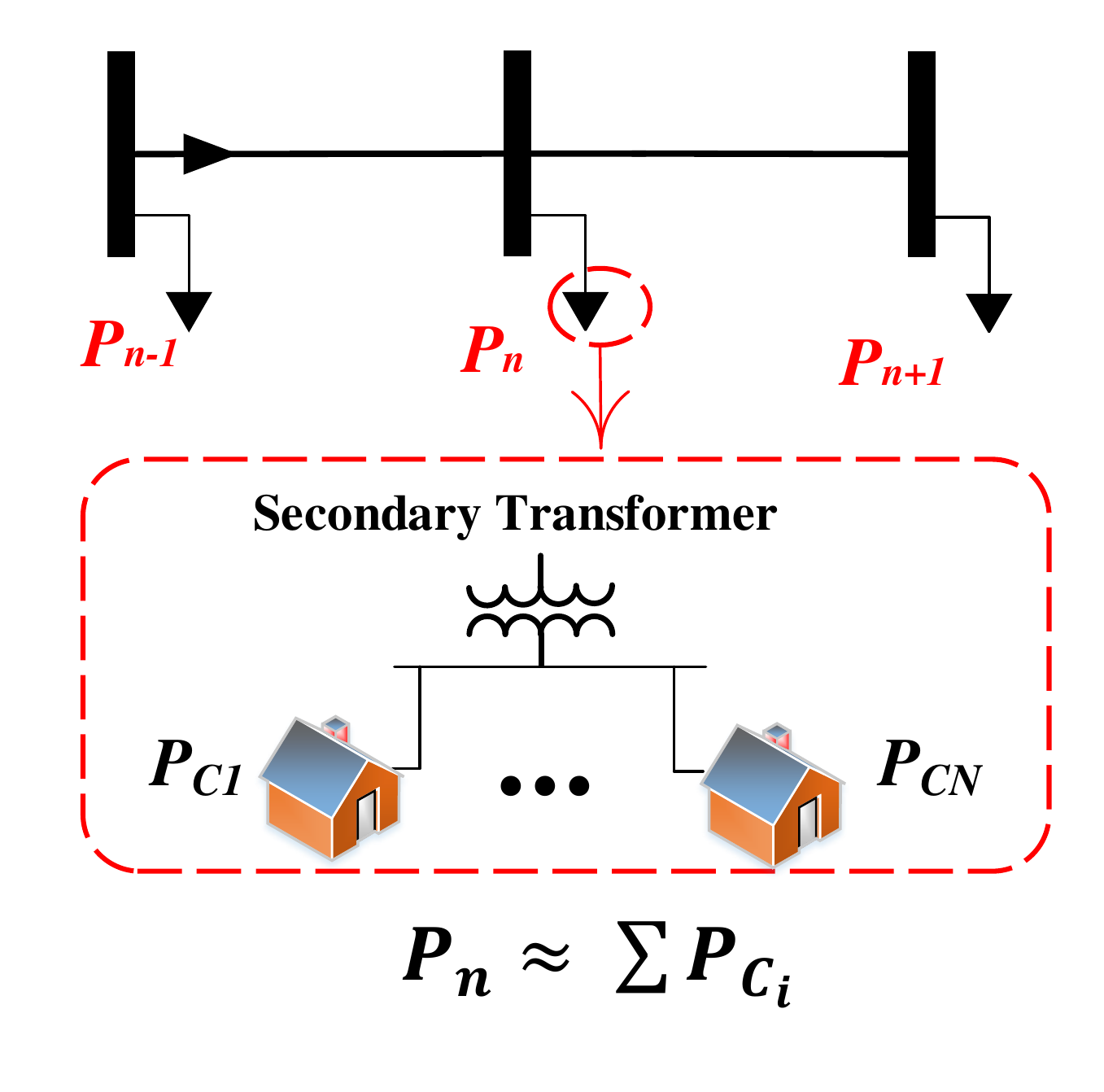}
	\caption{Data aggregation process for data privacy preservation.}
	\label{fig:aggreg}
\end{figure}



\section{Description of the Time-Series Test System}\label{data}
\subsection{System Description}
The time-series test system is a radial distribution system consisting of three feeders which are supplied by a 69 kV substation, as shown in Fig. \ref{fig:overall}. This test system is a real distribution grid located in the Midwest U.S. The real system belongs to a municipal utility and is a fully observable network with smart meters installed at all customers. The test system has 240 primary network nodes and 23 miles of primary feeder conductor. Notice that customers are connected to these primary network nodes via secondary distribution transformers, as can be seen in Fig. \ref{fig:aggreg}.

\subsection{Time-series Data Description}
In this test system, the time-series data of each node is directly obtained from customers' SM measurements. The data ranges from January 2017 to December 2017. The available SM data contains hourly energy consumption (kWh) of $1120$ customers. To perform time-series power flow analysis, the hourly average kW demand is approximated by the hourly energy consumption based on the assumption that the customer demand is constant in each one-hour time interval \cite{WHK2006}. To determine the reactive power, the power factor of each customer is randomly picked in the range of 0.9 to 0.95. A statistical approach is utilized to remove grossly erroneous and missing data samples \cite{kaveh2019}. Unlike the short-term demand data, our one-year period load data captures seasonal variations of customer behaviors. Fig. \ref{fig:season} shows one week load patterns at a selected primary node in different seasons.

\subsection{Load Tap Changing Substation Transformer}
The three feeders in the presented test system are supplied by a 69/13.8 kV step-down three-phase substation transformer with an on-load tap changing mechanism. The kVA rating and connection of the substation transformer are 10,000 kVA and delta-wye, respectively. The primary winding of the substation transformer is connected to a sub-transmission system which is equivalent to a swing bus. The equivalent short circuit impedance of the sub-transmission system are also provided. Accordingly, the positive-sequence resistance, positive-sequence reactance, zero-sequence resistance, and zero-sequence reactance are 4.5426 $\Omega$, 10.5274 $\Omega$, 7.3655 $\Omega$ and 24.5046 $\Omega$, respectively.

The tap-changing mechanism consists of three independent single-phase tap changers. The voltage ratings, kVA ratings, voltage settings, bandwidth, maximum voltage limit, minimum voltage limit, and the number of steps of each tap changer are 7.9674 kV, 3500 kVA, 123 V, 2 V, 129 V, 110 V, and 16, respectively. The three voltage tap changers are connected in wye-wye. The three-phase voltage of Bus 1 in the test system is monitored and controlled by the tap changers to implement voltage regulation.

\subsection{Secondary Distribution Transformers}
In the test system, the secondary distribution transformers consist of three-phase transformers and single-phase center-tapped transformers with different types, as shown in Table \ref{tbl:3_ph_distri_transf}. Most of the three-phase secondary distribution transformers are used for serving commercial customers with secondary nominal phase-to-phase voltage of 240 V, and most of the single-phase secondary distribution transformers are used for serving residential customers with secondary nominal voltage of 120/240 V.
\begin{table}[tbph]
\begin{center}
	\renewcommand{\arraystretch}{1.5}
	\caption{Parameters of Distribution Transformers}\label{tbl:3_ph_distri_transf}
	\label{cap}
	\begin{tabular}{ccccc} \hline
		\begin{tabular}[c]{@{}c@{}}  \textbf{Number of Phases} \\ \end{tabular} & \begin{tabular}[c]{@{}c@{}}  \textbf{Capacity}  \end{tabular} & $\mathbf{R}$ (\%)  & $\mathbf{X}$ (\%)  \\
	    \hline
        3 phases & 45 kVA  & 2.52  &  1.73  \\
        3 phases & 75 kVA & 2.27  &  1.91  \\
        3 phases & 112.5 kVA & 2.43 & 3.87 \\
        3 phases & 225 kVA & 1.15  &  5.5  \\
        3 phases & 300 kVA & 1.8  &  4.5 \\
        3 phases & 500 kVA & 1.6  &  5.9 \\
        1 phase & 15 kVA & 1.6  &  2.02  \\
        1 phase & 25 kVA & 1.4  &  2.3  \\
        1 phase & 37.5 kVA & 3.6  &  2.7  \\
        1 phase & 50 kVA & 3.1  &  2.8  \\
        1 phase & 100 kVA & 2.12  &  3.55  \\

    	\hline
	\end{tabular}
\end{center}
\end{table}
\begin{table}[tbph]
\begin{center}
	\renewcommand{\arraystretch}{1.2}
	\caption{Parameters of Conductors}\label{tbl:patameters_conductor}
	\label{conductor}
	\begin{tabular}{cccccc} \hline
		
	\textbf{Size}  &   \textbf{Material}    & \begin{tabular}[c]{@{}c@{}}  \textbf{Resistance} \\ \textbf{($\mathbf\Omega$/mile)} \end{tabular} & \begin{tabular}[c]{@{}c@{}}  \textbf{Diameter} \\ \textbf{(inch)} \end{tabular}  &  \begin{tabular}[c]{@{}c@{}}  \textbf{GMR} \\ \textbf{(feet)} \end{tabular}   &  \begin{tabular}[c]{@{}c@{}}  \textbf{Capacity} \\ \textbf{(A)} \end{tabular}\\
		
	    \hline
        4/0  & ACSR & 0.592 & 0.563 & 0.00814 & 340\\
	
        1/0  & ACSR & 1.12 & 0.355 & 0.00446 & 230\\

        4  & ACSR & 2.55 &  0.257 & 0.00452 & 140 \\

        2  & ACSR & 1.65 & 0.316 & 0.00504 & 180\\

        6  & CU & 2.41 &  0.201 & 0.00568 & 130 \\

        2  & CU & 0.87 &  0.3 & 0.0083 & 200 \\

        4/0  & AA & 0.554 &  0.512 & 0.0167 & 326\\
	
        1/0  & AA & 1.114 &  0.362 & 0.0111 & 228\\

    	\hline

	\end{tabular}
\end{center}
\end{table}
\begin{figure}[tbhp]
	\centering
	\includegraphics[width=3.5in]{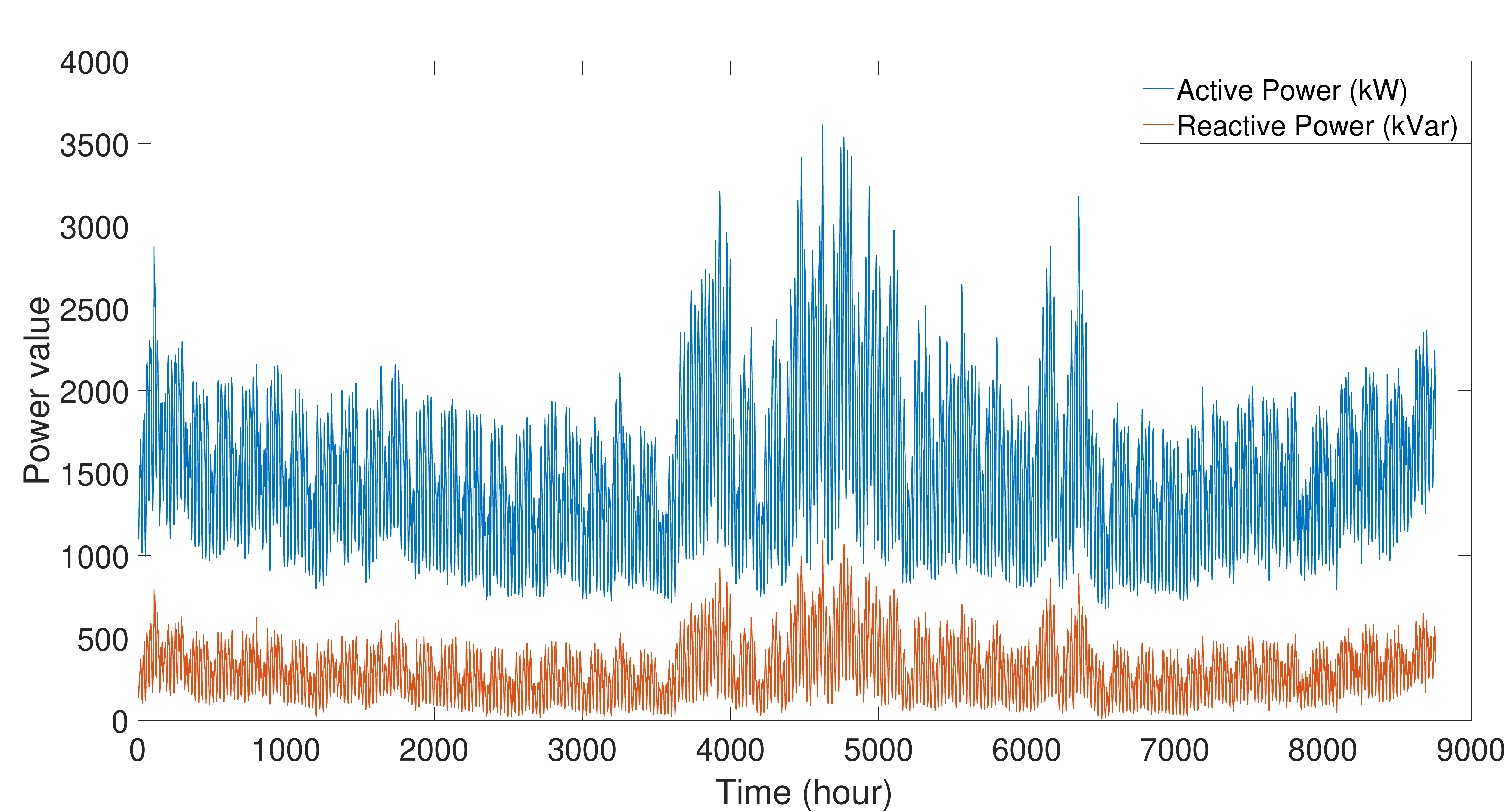}
	\caption{One-year active and reactive power consumption at the substation transformer.}
	\label{fig:apparent_power}
\end{figure}
\begin{figure}[tbhp]
\centering
\subfloat[Phase A\label{sfig:V_feederA}]{
\includegraphics[width=0.95\linewidth]{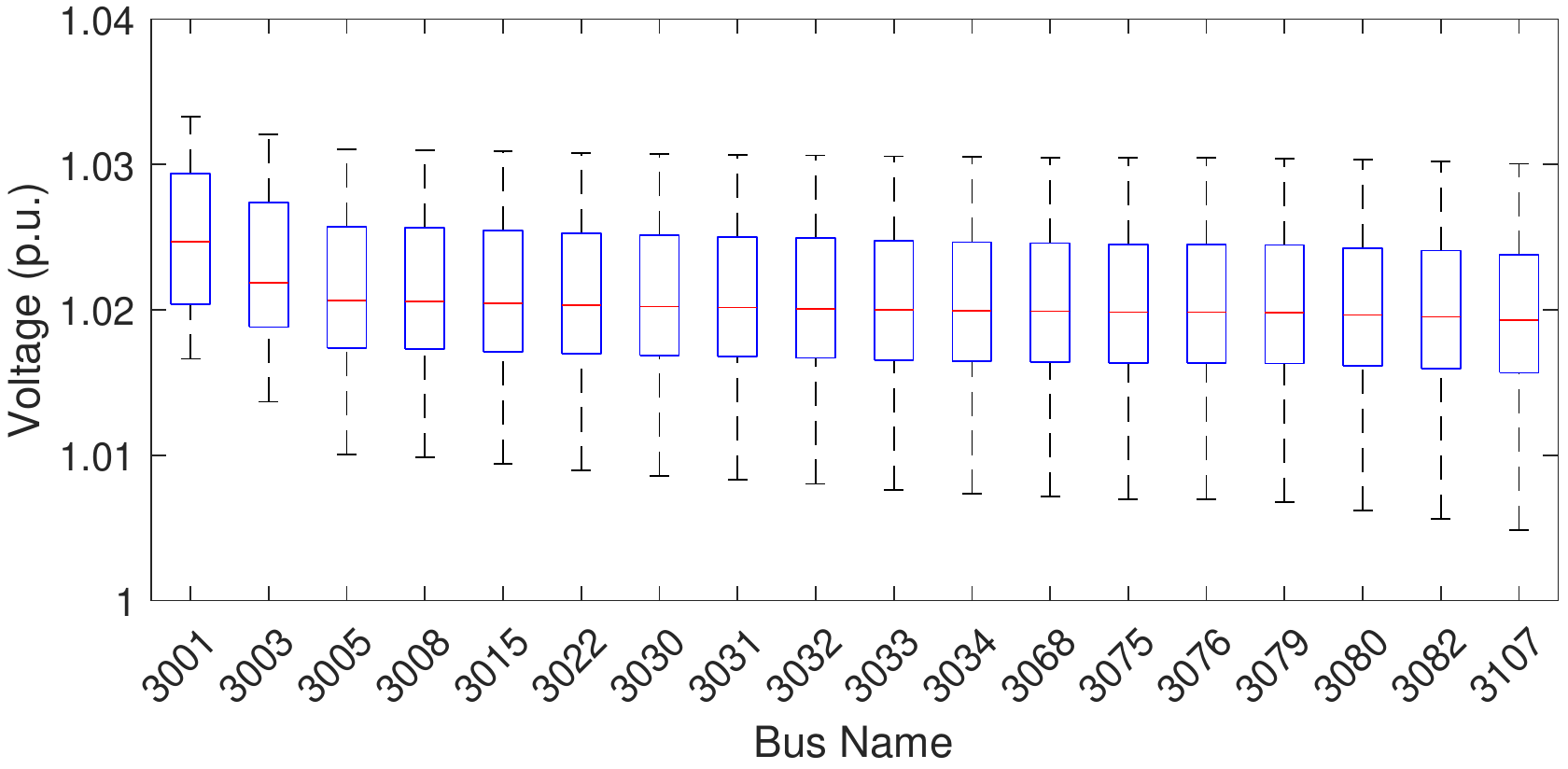}
}
\hfill
\subfloat[Phase B\label{sfig:V_feederB}]{
\includegraphics[width=0.95\linewidth]{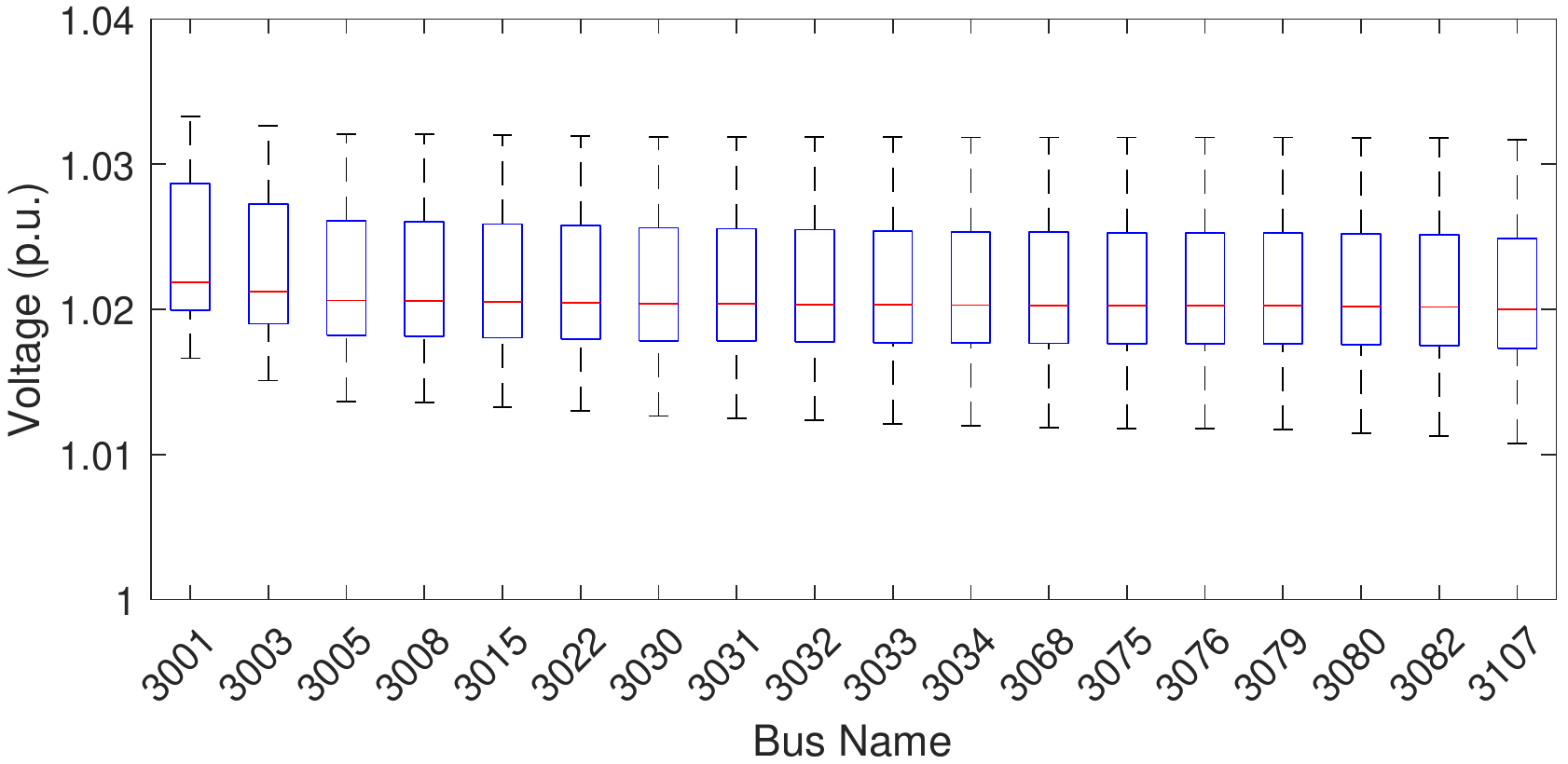}
}
\hfill
\subfloat[Phase C\label{sfig:V_feederC}]{
\includegraphics[width=0.95\linewidth]{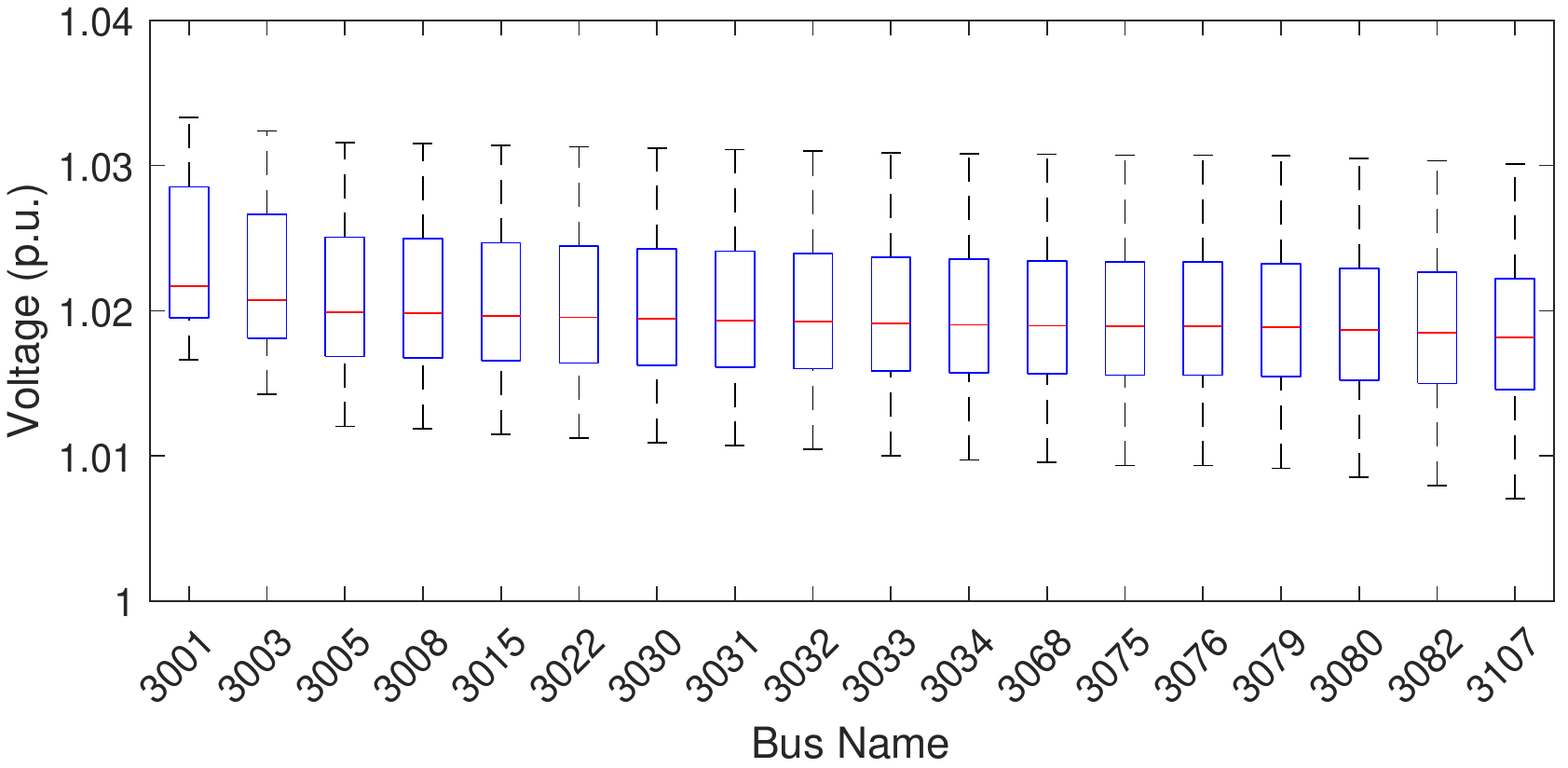}
}
\caption{Distributions of one-year nodal voltages of Feeder C.}
\label{fig:V_feeders}
\end{figure}


\subsection{Line}
The test system consists of overhead lines and underground cables with a variety of phasing configurations. The conductor and construction information is obtained from the real system to calculate series impedance and shunt capacitance. The conductor parameters are shown in Table \ref{tbl:patameters_conductor}, where, ACSR denotes Aluminum Conductor Steel Reinforced, AA denotes All Aluminum, and CU denotes Copper. The series impedance and shunt capacitance of conductors are calculated as follows \cite{WHK2006}:

\subsubsection{Overhead Lines}
For conductor $i$ and conductor $j$ of an overhead line, the self impedance of conductor $i$ and mutual impedance between conductors $i$ and $j$ are calculated by the modified Carson's equations \cite{WHK2006}:
\begin{equation}
\label{eq:self}
\hat{z}_{ii}=r_i+0.09530+j0.12134(ln\frac{1}{GMR_i}+7.93402)
\end{equation}
\begin{equation}
\label{eq:mut}
\hat{z}_{ij}=0.09530+j0.12134(ln\frac{1}{D_{ij}}+7.93402)
\end{equation}
where, $r_i$ is the resistance of conductor $i$, $GMR_i$ is the geometric mean radius of conductor $i$, and $D_{ij}$ is the distance between conductors $i$ and $j$. Then, the primitive impedance matrix, $\mathbf{\hat{z}}_{primitive}$, can be built, and the phase impedance matrix, $\mathbf{z}_{abc}$, can be obtained by employing Kron reduction to reduce the dimensions of primitive matrix, as follows \cite{WHK2006}:
\begin{equation}
\label{eq:primitive}
\mathbf{\hat{z}}_{primitive}=
\left[
\begin{array}{ccc}
\mathbf{\hat{z}}_{ij} & \mathbf{\hat{z}}_{in} \\
\mathbf{\hat{z}}_{nj} & \mathbf{\hat{z}}_{nn}
\end{array}
\right]
\end{equation}
\begin{equation}
\label{eq:kron}
\mathbf{z}_{abc}=\mathbf{\hat{z}}_{ij}-\mathbf{\hat{z}}_{in}\cdot \mathbf{\hat{z}}_{nn}^{-1}\cdot \mathbf{\hat{z}}_{nj}
\end{equation}
Since the shunt admittance of an overhead line is small, in this test system, the shunt capacitance is ignored \cite{WHK2006}.
\begin{figure}[tbhp]
\centering
\subfloat[Phase A\label{sfig:tap_V_A}]{
\includegraphics[width=0.98\linewidth]{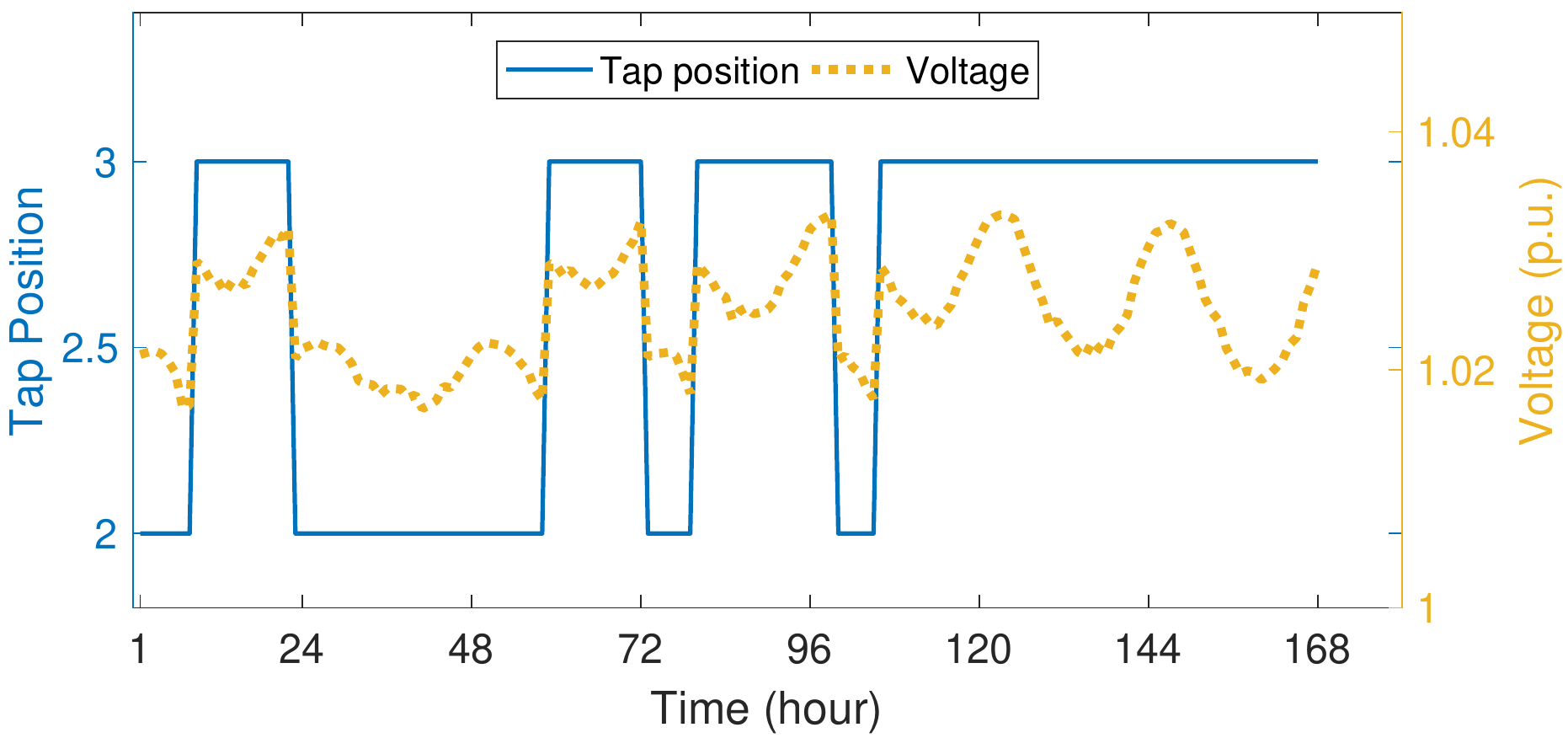}
}
\hfill
\subfloat[Phase B\label{sfig:tap_V_B}]{
\includegraphics[width=0.98\linewidth]{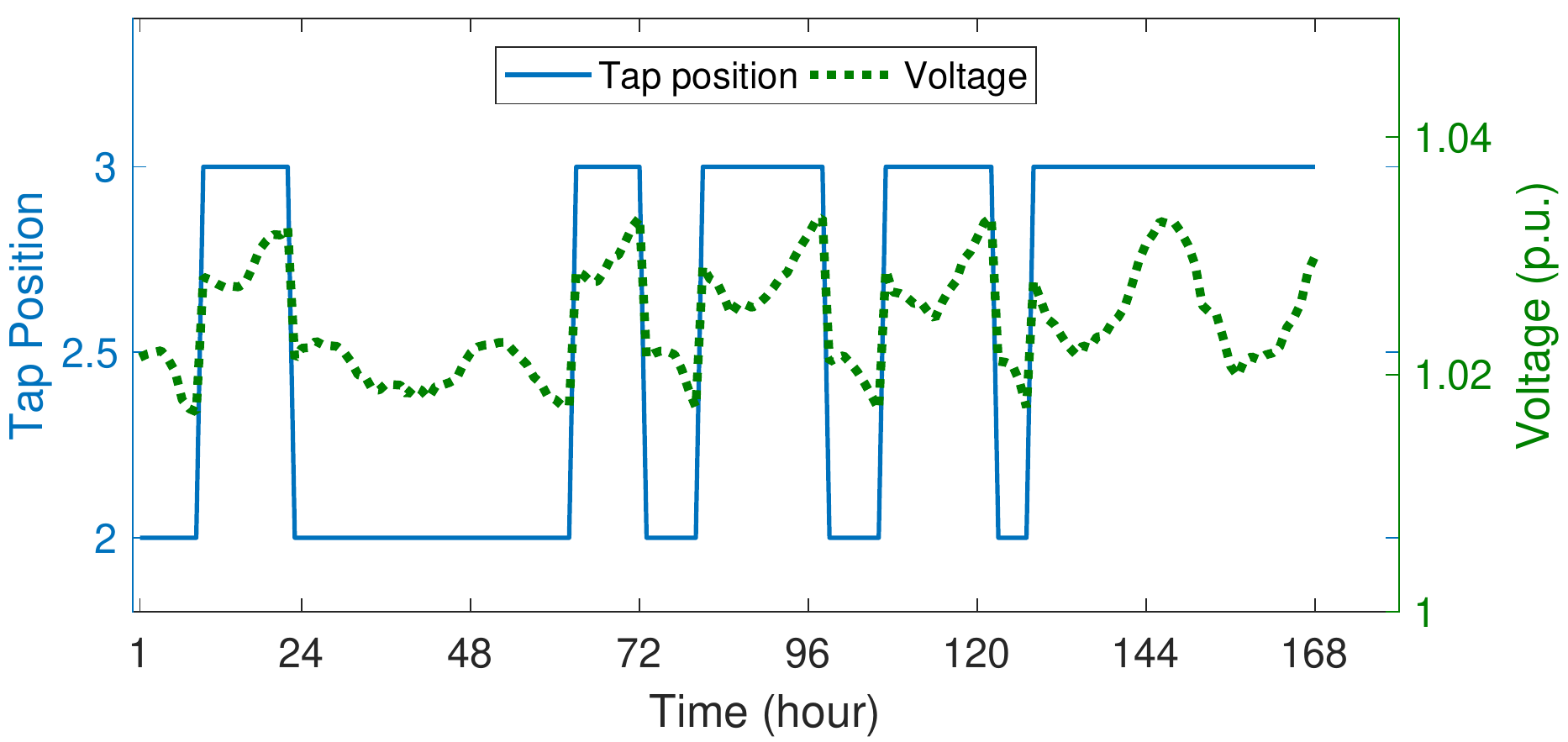}
}
\hfill
\subfloat[Phase C\label{sfig:tap_V_C}]{
\includegraphics[width=0.98\linewidth]{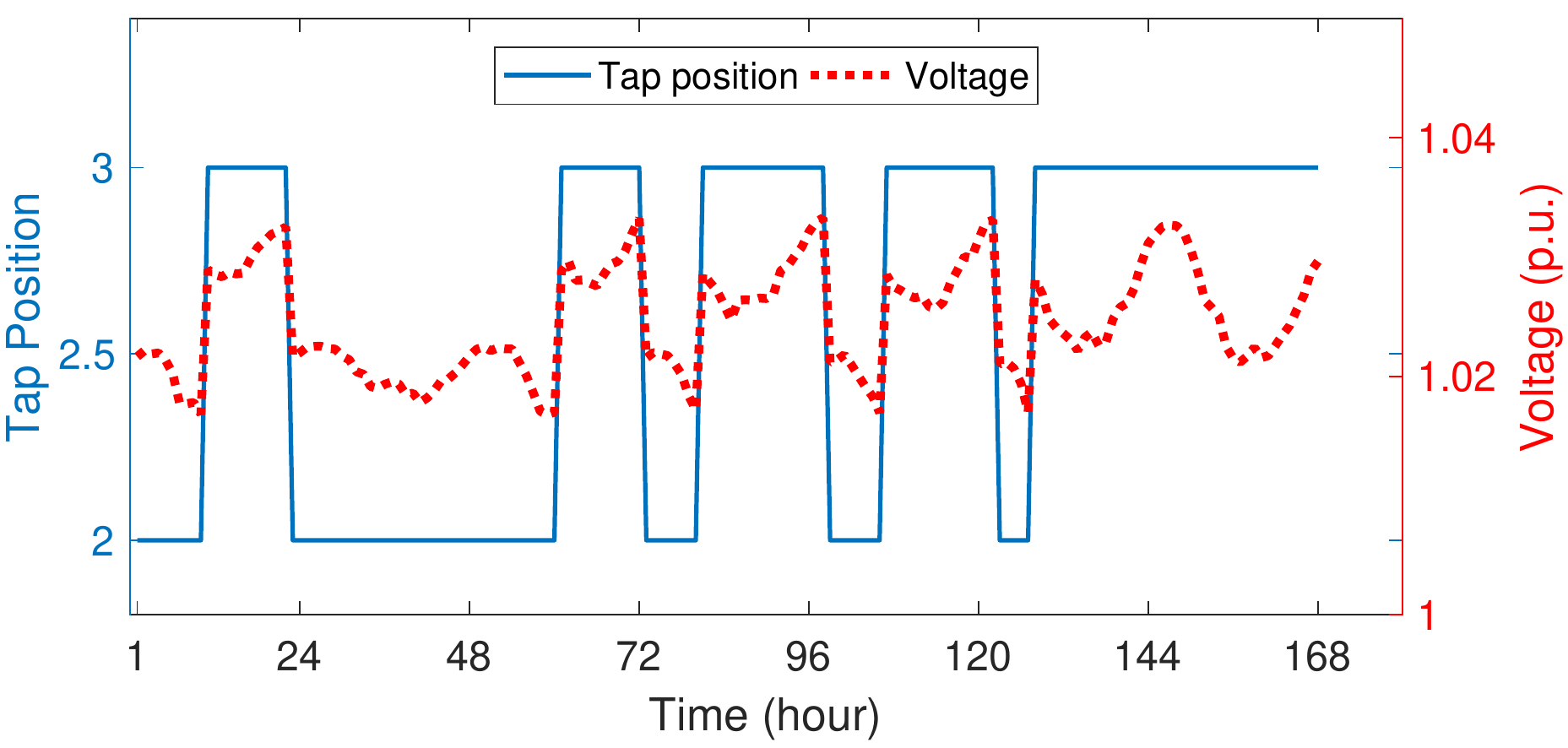}
}
\caption{Tap positions and voltage magnitudes at Bus 1, from July 1st to July 7th, 2017.}
\label{fig:tap_V}
\end{figure}
\begin{figure}[tbhp]
	\centering
	\includegraphics[width=3.5in]{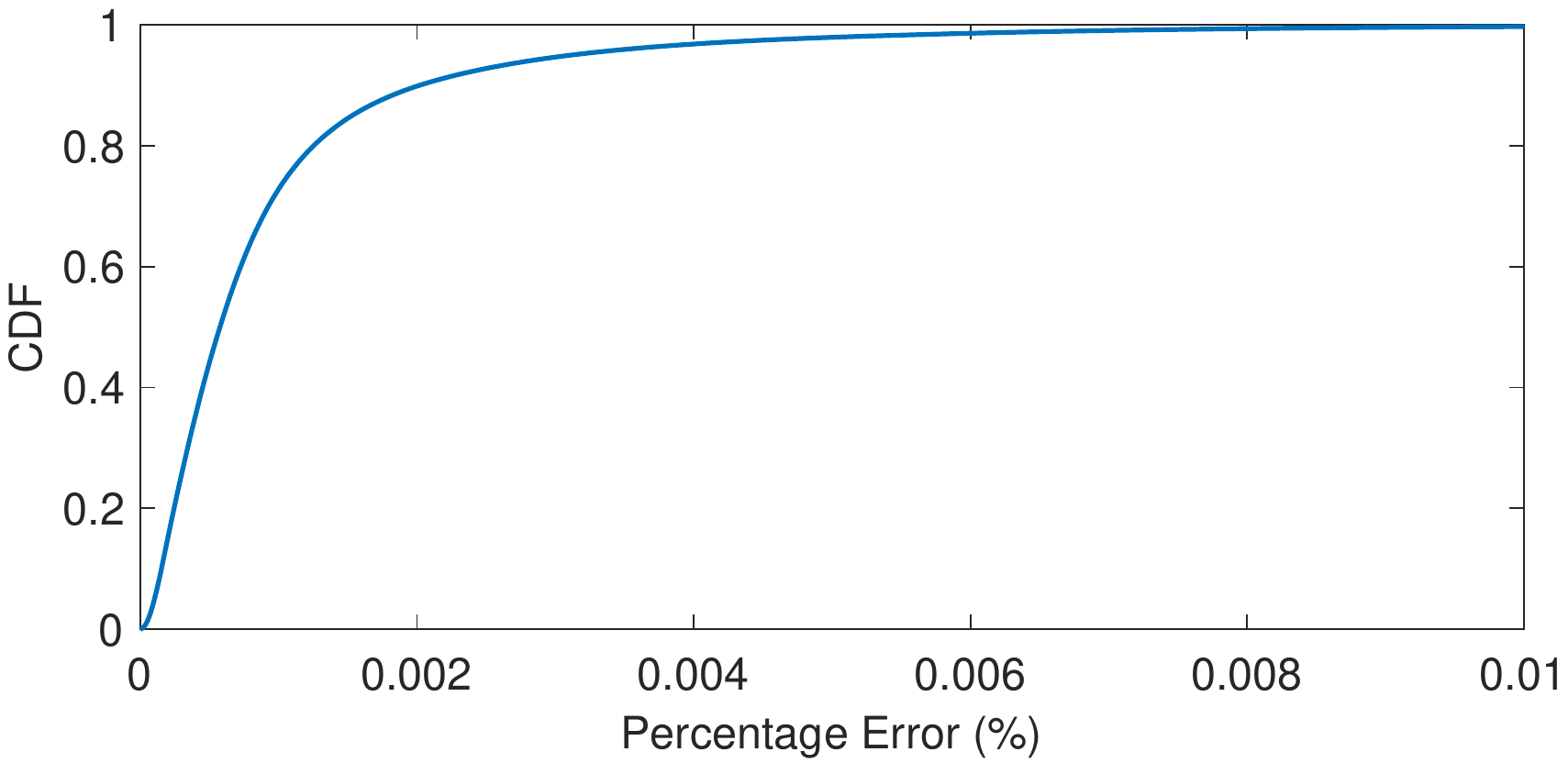}
	\caption{Cumulative distribution function of the power flow solution error.}
	\label{fig:errors}
\end{figure}

\subsubsection{Underground Cables}
The Equation \eqref{eq:self}-\eqref{eq:kron} are also used for calculating series impedance of underground cables \cite{WHK2006}. Compared with the overhead lines, the shunt capacitance of underground cables cannot be ignored and is calculated as follows:
\begin{equation}
\label{eq:capacitance}
C_{pg}=\frac{2\pi\varepsilon}{ln(R_b/RD_c)-(1/k)ln(k \cdot RD_s/R_b)}
\end{equation}
where, $\varepsilon$ is the permittivity of medium, $R_b$ represents the radius of a circle passing through the center of neutral strands, $RD_c$ is the radius of phase conductor, $k$ is the number of strands, and $RD_s$ is the radius of strand conductor.

\subsection{Shunt Capacitor Bank}
The test system has two shunt capacitor banks for voltage regulation, which are located at Feeder B and Feeder C, respectively. The kVAr rating and the connection of the two capacitor banks are 50 and grounded-wye, respectively. This utility has a strategy to switch on capacitor banks in normal operation to provide reactive power support.

\subsection{Circuit Breaker}
In the test system, the three feeders have a total of 9 circuit breakers for system protection and reconfiguration, six of which are normally-closed and the remaining three are normally-open.

\section{A Numerical Example of the test system}\label{result}
The time-series simulation of the test system is performed using the OpenDSS simulation program, which is a commonly-used open-source solver \cite{opendss}. Matlab-OpenDSS COM interface is employed to run time-series power flow over a one-year period. Meanwhile, the power flow results that include bus voltages and line currents are collected through the interface.

Voltage level is a critical concern for utilities, therefore, it is of importance to analyze the range of bus voltages in the power flow solutions. Fig. \ref{fig:V_feeders} shows the statistical results of the primary main feeder buses' voltages of Feeder C over a year. It can be seen that the voltage in this test system is within the range of 1.005 $p.u.$ to 1.035 $p.u.$, which satisfies the voltage quality requirement.

Fig. \ref{fig:tap_V} shows the tap positions of the load tap changing transformer and voltage profiles of Bus 1 during one peak demand week from July 1st to July 7th. Fig. \ref{sfig:tap_V_A}, Fig. \ref{sfig:tap_V_B} and Fig. \ref{sfig:tap_V_C} correspond to phase A, phase B and phase C, respectively. It can be seen that once the voltage drops below the voltage setting, the regulator will increase the tap position from 2 to 3, and consequently, the voltage increases.

To check the convergence of the OpenDss simulation model, error checking is conducted on the time-series power flow solution. The calculated power of each bus is obtained based on the bus voltages and line currents from the power flow solution. Then, the error is obtained by evaluating the difference between the calculated power and the expected power. Fig. \ref{fig:errors} shows the cumulative distribution function (CDF) of the percentage error, where $95\%$ of the error samples are less than $(3\times10^{-3})\%$.

This test system has been used in the literature \cite{kaveh2019}-\cite{cascade} to develop and verify machine learning-based algorithms for load inference and distribution system state estimation. The OpenDSS model of this real distribution grid can be downloaded from \cite{zhaoyu_wang}.

\section{Conclusion}\label{conclusion}
This paper presents a distribution test system based on a real distribution grid in Midwest U.S. We provide the real network topologies and electric equipment parameters. Compared to existing test systems, this system is unique as we have released the associated one-year real measurement data from smart meters. In addition, the OpenDSS model of this test system is available on our website. The test system provides researchers an opportunity to validate and demonstrate their theoretical work using a real distribution grid model with field measurements.

\ifCLASSOPTIONcaptionsoff
  \newpage
\fi

\bibstyle{IEEEtran}
\bibliography{IEEEabrv,./bibtex/bib/IEEEexample}

\end{document}